\documentclass[aps,prb,twocolumn,groupedaddress]{revtex4}
\usepackage{graphicx}
\newcommand{\ket}[1]{| #1 \rangle}
\newcommand{\bra}[1]{\langle #1 |}

\newcommand{\ex}[1]{\langle #1 \rangle}

\newcommand{\beq}{\begin{eqnarray}}
\newcommand{\eeq}{\end{eqnarray}}

\begin{document}

\title {Electron spin manipulation and resonator readout in a double quantum dot nano-electromechanical system}
\author{N. Lambert}
\affiliation{Department of Basic Science, University of Tokyo, Komaba, Meguro-ku, Tokyo 153-8902, Japan}

\email{lambert@asone.c.u-tokyo.ac.jp}
\author{I. Mahboob}
\affiliation{NTT Basic Research Laboratories, NTT Corporation, Atsugi-shi, Kanagawa 243-0198, Japan}

\author{M. Pioro-Ladri\`{e}re}
\affiliation{Quantum Spin Information Project, ICORP, Japan Science and Technology Agency, Atsugi-shi, Kanagawa 243-0198, Japan}

\author{Y. Tokura}


\affiliation{NTT Basic Research Laboratories, NTT Corporation, Atsugi-shi, Kanagawa 243-0198, Japan}
\affiliation{Quantum Spin Information Project, ICORP, Japan Science and Technology Agency, Atsugi-shi, Kanagawa 243-0198, Japan}

\author{S. Tarucha}
\affiliation{Department of Applied Physics, University of Tokyo, Hongo, Bunkyo-ku, Tokyo, 113-0033, Japan}
\affiliation{Quantum Spin Information Project, ICORP, Japan Science and Technology Agency, Atsugi-shi, Kanagawa 243-0198, Japan}

\author{H. Yamaguchi}


\affiliation{NTT Basic Research Laboratories, NTT Corporation, Atsugi-shi, Kanagawa 243-0198, Japan}
\affiliation{Department of Physics, Tohoku University, Sendai, Miyagi 980-8578, Japan}

\begin{abstract}
\noindent Magnetically coupling a nano-mechanical resonator to a
double quantum dot confining two electrons can enable the
manipulation of a single electron spin and the readout of the
resonator's natural frequency. When the Larmor frequency matches the
resonator frequency, the electron spin in one of the dots can be
selectively flipped by the magnetised resonator. By simultaneously
measuring the charge state of the two-electron double quantum dots,
this transition can be detected thus enabling the natural frequency
of the mechanical resonator to be determined.

\end{abstract}

\maketitle

\section{Introduction}

Nano-electromechanical systems (NEMSs) enable the coupling of
electronic and mechanical degrees of freedom \cite{Craig, Cho,
Roukes1} and provide an arena in which quantum mechanical behaviour
can be identified on macroscopic scales
\cite{Craig, Cho, Roukes1, Cleland, Schwab, Schwab2}. A key
requirement for the realization of quantum electromechanical systems
is the efficient transduction of the miniscule displacements of the
mechanical element. Meanwhile theoretical studies have proposed
coupling NEMSs to quantum two-level systems realised in a variety of
physical systems \cite{Gell,xue2,Carr,Hu}. The coupled system can
potentially enable the implementation of protocols similar to those
realised in cavity QED and on-chip circuit QED with superconducting
qubits and transmission lines\cite{Blais}. Furthermore, manipulating
a two-level system via NEMSs has the advantage that a change in its
state can be used to distinguish between resonance and off resonance
conditions of the mechanical element with great sensitivity.

\begin{figure}[!hbt]
\includegraphics[width=1\columnwidth]{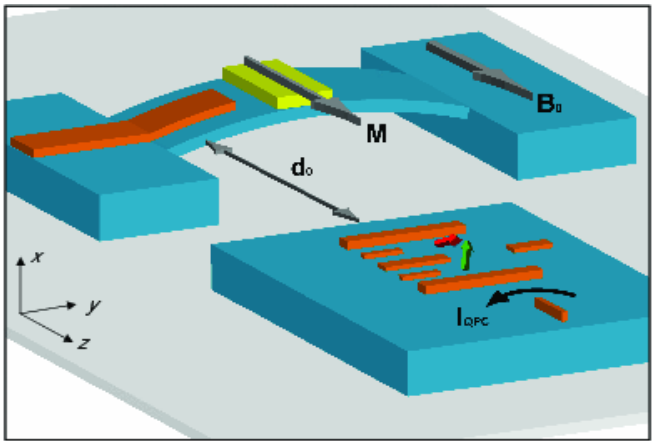}
\vspace{0.0cm} \caption{\label{fig1x} (colour on-line) A schematic
of the proposed coupled mechanical resonator double QD system. The
mechanical oscillator is a suspended doubly clamped beam located a
distance, $d_0$, from the mesa on which the double QDs are defined
via electrostatic gates (orange). The beam is actuated to resonance
via the piezoelectric effect between the top gate, located at where
the beam is clamped (orange), and the shallow buried 2DEG. A
micro-magnet (yellow) of magnetisation, $M$, is incorporated onto
the beam oscillator and moves in unison with the out of plane beam
motion. The mechanical oscillator couples to the confined electron
spins (red and green arrows) in the doubles QDs via the
micro-magnet's stray magnetic field where the degree of coupling is
controlled by the tuning field, $B_0$. The quantum point contact
current, $I_{QPC}$, is used to measure the charge state of the QDs.}
\end{figure}

In this work we present a proposal for coupling a NEM system to the
spin of an electron in a double quantum dot (QD).  Quantum dots are
artificial structures which enable the confinement of a few or even
one electron \cite{Ashoori96, Tarucha96, Ciorga00}. In this regime,
the spin of the individual electron forms a natural two-level system
and provides great potential for the realization of a solid-state
qubit \cite{Loss98}. However, for a QD based spin qubit to be
realized requires precise control and manipulation of the electron
spins. Recently, coherent spin manipulation in a double quantum dot
was demonstrated by rapid electrical control of the exchange
interaction \cite{Petta05Sci} and by generating a local electron
spin resonance (ESR) magnetic field \cite{Koppens06}.

Here we describe how to exploit the versatility of a double quantum
dot in order to realize an electron spin in a well defined state
which is then coupled and controllably manipulated by the
nano-mechanical resonator. Previously, coupling spin to a mechanical
resonator has been both theoretically and experimentally considered
\cite{xue, Rugar}; but the motivation of these studies was spin
detection rather than spin manipulation. In the present scheme, the
magnetised resonator selectively flips one of the two confined
electron spins in the double dots (DDs) where the ESR field is
produced by the oscillations of the magnetised resonator. The
corresponding change in the two electron spin state is then detected
by spin blockade charge sensing at the DDs. Concurrently, this also
enables the resonator's natural frequency to be measured. As
demonstrated in refs. \cite{Petta05Sci, Koppens06}, spin-blockade
readout does not rely on large Zeeman fields as in the case of a
single quantum dot \cite{SQD} which is an important criterion for
frequency matching between the electron spin and the mechanical
resonator. Furthermore, the effects of the hyperfine interaction on the electron spin states
and the corresponding line width of the proposed measurement is also theoretically investigated.


\section{Device Geometry and Parameters}

A schematic of the proposed coupled nano-mechanical resonator DD
system is shown in fig.~\ref{fig1x}. A GaAs/AlGaAs nano-mechanical
beam of length 2 $\mu$m, width 500 nm and thickness 200 nm is
located a distance, $d_0$, from the mesa on which the DDs reside.
The natural frequency for a resonator of these dimensions is
expected to be more than 200 MHz and peak displacement, at the
centre of the beam, before the onset of non-linearity of $\sim$ 5 nm
assuming that the resonator can be driven hard enough with a
routinely achievable quality factor of 10$^4$.\cite{bi} The double
dots are defined using electrostatic gates on the surface of the
heterostructure where the 2DEG is located typically 100 nm below the
surface \cite{Elzerman03}.

\begin{figure}[!ht]
\includegraphics[scale=1.0]{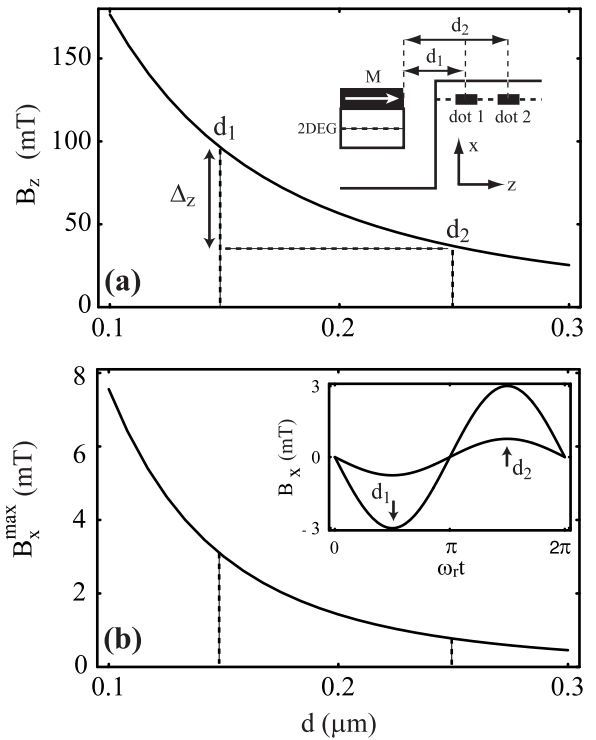}
\caption{\label{fig1} (a) The profile of the in-plane (Zeeman) magnetic field, $B_z$, near the nano-mechanical resonator with external tuning field, $B_0$ = 0 T where $d_1$ and $d_2$ mark the positions of the left and right quantum dots respectively. The inset shows the cross-section of the proposed system. The position of the micro-magnet is in the same plane as the 2DEG in the mesa on which the double QDs are located. This alignment is achieved by piezoelectrically flexing the beam with the application of DC voltage to the gate defined on the mechanical oscillator and is discussed in detail below. (b) The profile of the transverse magnetic field, $B_{x}^{\text{max}}$, whilst the resonator is at maximum deflection (5 nm). The inset shows the time dependence of the ESR field at the positions of the quantum dots whilst the mechanical element is at resonance.}
\end{figure}

The resonator is coupled to the DDs magnetically by placing a Co
micro-magnet on the resonator as depicted in fig.~\ref{fig1x}, with,
$\mu_0M = 1.8$ T, where $ \mu_0$ and $M$ are the permeability of
free space and magnetization of Co respectively. The Co micro-magnet
has a length 500 nm, width 200 nm and thickness 100 nm and is
engineered so that the magnetization vector remains parallel  to the
easy axis for small values ($\sim$ 50 mT) of the external tuning
field, $B_0$. The left and right dots are located a distance, $d_1$
= 150 nm and $d_2$ = 250 nm, from the mechanical resonator
respectively. Under these conditions, with the nano-mechanical beam at resonance, the total field
including the external tuning  field is of the form, $B = (B_0 +
B_{zn} + \delta B_{zn} \text{sin}(\omega_rt))\hat{z} -
B_{xn}\text{sin}(\omega_rt)\hat{x}$. The stray field of the Co micro-magnet has been calculated analogously to the method outlined in ref. \cite{Michel}.
The in-plane (Zeeman) magnetic field, $B_z$, arising from the above described magnetic element is
shown in fig.~\ref{fig1}a. At the QD locations, the in-plane
magnetic field at the left (right) dot is, $B_{z1(2)}$ = 94 mT (36
mT). The in-plane component of the magnetic field
exhibits almost no temporal dependence ($\delta B_{zn} \ll B_{zn}$)
even though the field originates from the magnetic element placed on
the resonator because the beam oscillation is much smaller than the thickness of the micro-magnet.
The proximity of the magnetic element to the DDs
produces a strong in-plane field gradient of order, $(B_{z1}-B_{z2})
/(d_1-d_2) \sim$ 0.6 T/$\mu$m which allows the nano-mechanical
oscillator to addressably couple to the electron spin in either
quantum dot as discussed below. The spatial dependence of the
transverse magnetic field, $B_{x}^{\text{max}}$, whilst the beam is
at maximum deflection is shown in fig.~\ref{fig1}b. In the inset of
fig.~\ref{fig1}b, the time dependence of the transverse magnetic
field, $B_{x}$, at the locations of the DDs whilst the beam is at
resonance is shown. The amplitude of this field at the location of
the left (right) dot is $\sim$ 3.0 mT (0.8 mT) which is more than
sufficient to flip the electron spin in either dot via ESR. As can
be seen from fig.~\ref{fig1}a and b, the parameter, $d_0$, is of
critical importance as this governs the degree to which the DDs
couple to the resonator via the magnetic element.

\subsection{Hamiltonian}

For two-electron QDs, the relevant spin states are the spin singlet
and triplet states of the (1,1) charge state ($|S\rangle,
|T_0\rangle, |T_+\rangle, |T_-\rangle$) and the spin singlet of the
(0,2) charge state ($|(0,2)S\rangle$) where the label ($m,n$) refers
to the number of electrons confined on the left and right dots
\cite{Taylor2005}. In the presence of finite interdot tunnelling,
$T$, the (1,1) and (0,2) singlets are coupled allowing electrons to
be moved between dots when the detuning parameter, $\epsilon$, is
varied (by pulsing the QD gate voltages). The energy, $\Delta_z =
g\mu_B (B_{z1}-B_{z2})/2$, couples the $|S\rangle$ and $|T_0\rangle$
states whilst $|T_+\rangle$ and $|T_-\rangle$ remain eigenstates
with Zeeman energy, $\pm E_Z = \pm g\mu_B (B_{z1}+B_{z2} + 2B_0)/2$,
where, $g$, is the electron g-factor ($|g|$ = 0.44 for GaAs) and,
$\mu_B$, is the Bohr magneton. For the stationary resonator, the
Hamiltonian describing the two electron states in the presence of
the above described in-plane in-homogeneous magnetic field in the
$\{|(0,2)S\rangle, |S\rangle, |T_0\rangle, |T_+\rangle,
|T_-\rangle\}$ basis is

\begin{equation}
H =
\left[ \begin{array}{ccccc}
-\epsilon & \sqrt{2}T & 0 & 0 & 0 \\
\sqrt{2}T & 0 & -\Delta_z & 0 & 0 \\
0 & -\Delta_z & 0 & 0 & 0 \\
0 & 0 & 0 & -E_Z & 0\\
0 & 0 & 0 & 0 & E_Z\\
\end{array}\right].
\label{hamil}
\end{equation}

\begin{figure}[!ht]
\includegraphics[scale=1.0]{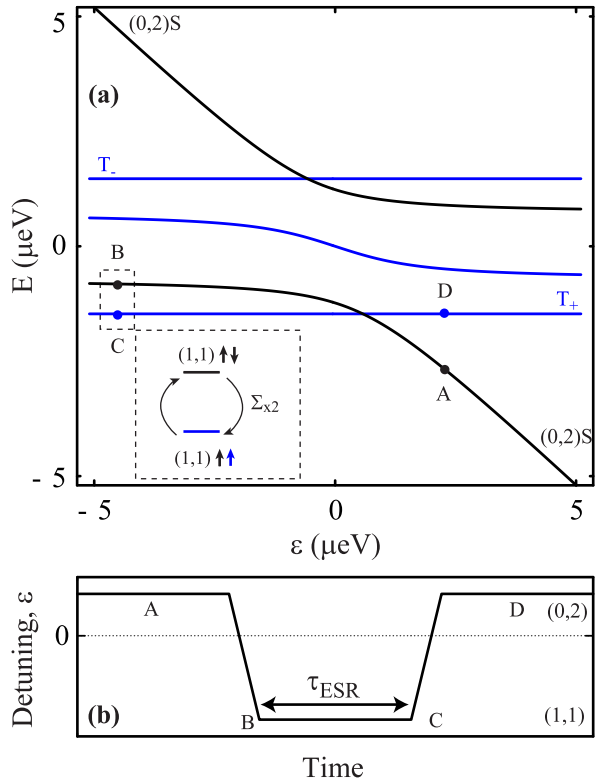}
\vspace{-1.0cm} \caption{\label{fig2} (a) (colour on-line) The
two-electron spin states as a function of detuning energy with
interdot tunnelling, $\sqrt{2}T$ = 1 $\mu$eV with the external
tuning field set to zero. At large negative detuning, the transverse
ESR field ($B_{x2}$) resonantly couples the
$\left|\uparrow\downarrow\right\rangle$ and $|T_+\rangle$ states
(blue line) when the condition, $g\mu_{B}(B_{z2} + B_0) =
\hbar\omega_r$, is satisfied. (b) The pulse sequence consisting of
preparation in the $|(0,2)S\rangle$ (A), resonant coupling of the
$\left|\uparrow\downarrow\right\rangle$ and $|T_+\rangle$ states for
time, $\tau_{ESR}$, (B$\rightarrow$C) and readout (D) steps. For,
$\tau_{ESR} = \tau_{\pi}$, the final state at point (D) is the
triplet $\left|T_{+}\right\rangle$.}
\end{figure}

In fig.~\ref{fig2}a, the eigenenergies of $H$ are plotted as the
detuning is swept between the (1,1) and (0,2) charge states across
the degeneracy point ($\epsilon = 0$). For large positive detuning,
the ground state is $|(0,2)S\rangle$. Close to $\epsilon = 0$, $T$
and $\Delta_z$ mix the $|(0,2)S\rangle$, $|S\rangle$ and
$|T_0\rangle$ states. For large negative $\epsilon$,
$|(0,2)S\rangle$ uncouples from $|S\rangle$ and $|T_0\rangle$ to
become the highest energy state. In the absence of a field gradient
($\Delta_z = 0$), $|S\rangle$ and $|T_0\rangle$ would also be
eigenstates with splitting equal to the exchange energy, $J \approx
T^2/\epsilon$. However, because $\Delta_z \gg J$, the actual
eigenstates are the superpositions, ($(|S\rangle \pm
|T_0\rangle)/\sqrt{2}$), which correspond to the states
$\left|\uparrow\downarrow\right\rangle$ (spin up and spin down at
the left and right QDs respectively) and
$\left|\downarrow\uparrow\right\rangle$ respectively. Moreover, the
field gradient in the regime of large negative detuning results in
asymmetric separation of $|T_+\rangle$ and $|T_-\rangle$ states
about $\left|\uparrow\downarrow\right\rangle$ where the energetic
separation between the $\left|\uparrow\downarrow\right\rangle$ and
$|T_+\rangle = \left|\uparrow\uparrow\right\rangle$ states is,
$\Sigma_{z2} = g\mu_{B}(B_0 + B_{z2})$, the Zeeman energy of the
right spin.

\section{Spin Isolation and ESR}

\begin{figure}[!ht]
\includegraphics[width=1\columnwidth]{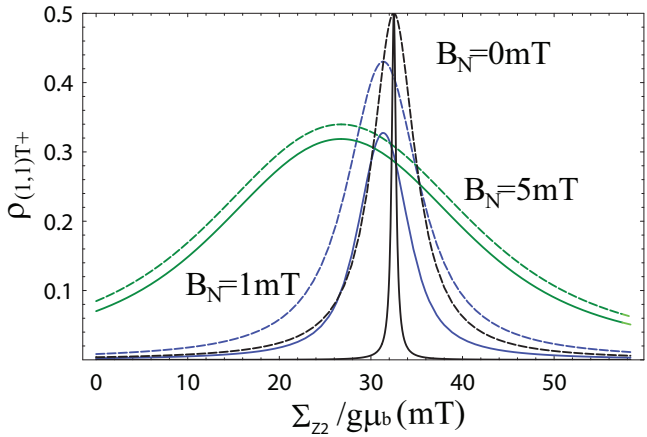} 
\vspace{-.5cm} \caption{\label{fig3} The expected QPC measurement
for the occupation of the $\ket{1,1}T_+$ state as a function of the
field, $\Sigma_{z2}/g\mu_B$, for long ESR mixing time, $\tau_{ESR}$,
when the right QD is resonantly coupled to the mechanical element.
The resolution of the ESR peak is reduced and broadened by the
increasingly strong nuclear field. The range of fields presented
here are $\ex{B_{N}}=$ 0, 1, 5 mT, with the expected hyperfine field
in GaAs  being 1mT.  The solid lines represent $Q=10^4$ while the
dashed lines represent $Q=10^5$, illustrating the inverse dependance
of the QPC line shape on the quality factor of the resonator. Here
the incoherent $\gamma$ process is negligibly small. }
\end{figure}

 The nano-mechanical resonator can be used to manipulate the
electron spin states in the DDs as follows. The resonator is excited
at its natural frequency corresponding to an energy,
$\hbar\omega_r$, the equivalent magnetic field, $\hbar\omega_r /
g\mu_B$ is designed to be much greater than GaAs hyperfine magnetic
field, $B_{N} \sim$ 1 mT. The in-plane tuning magnetic field is then
used to adjust the Zeeman spin splitting to couple the right QD and
the mechanical element by setting $\Sigma_{z2} = \hbar\omega_r$.
That is, a tuning field corresponding to the energetic difference
between the Zeeman spin splitting at the right QD and the mechanical
oscillator results in resonant coupling. For a 200 MHz resonator
with the above values for the in-plane magnetic field, the tuning
field will be -3.5 mT. With the mechanical element at resonance, the
$|(0,2)S\rangle$ state is prepared (A) which is then separated into
the $\left|\uparrow\downarrow\right\rangle$ state by the in-plane
magnetic field gradient between the two QDs and by using rapid
adiabatic passage to change $\epsilon$ from a positive to a negative
value (A$\rightarrow$B) \cite{Petta05Sci}. With the right QD
resonantly coupled to the resonator, the transverse ESR magnetic
field couples only the $\left|\uparrow\downarrow\right\rangle$ and
$|T_+\rangle$ states at a rate proportional to
$\Sigma_{x2}=g\mu_{B}B_{x2}$. The left spin is unaffected by the ESR
field due to the large difference in Zeeman field between the QDs.
Assuming an infinite quality factor, the Hamiltonian in these
conditions is \beq H_{ESR}= -\Sigma_{z2} \sigma_{z2}/2 +
\Sigma_{x2}(\omega_r,\omega_d, Q) \text{sin}(\omega_dt)
\sigma_{x2}/2, \eeq where $\sigma_{n2}=(\sigma_{x2}, \sigma_{y2},
\sigma_{z2})$ are the Pauli matrices of the spin located in the
right QD, $\omega_r$ is the natural frequency of the resonator, and
$\omega_d$ is the driving frequency of the resonator. The magnitude
of the $\Sigma_{x2}$ field changes as a function of $Q$, the quality
factor, and $\omega_d$, according to its proportionality to the
displacement of the resonator \beq \Sigma_{x2}(\omega_r,\omega_d, Q)
=\frac{C}{\sqrt{(\omega_r^2 - \omega_d^2)^2 +
4\omega_r^2\omega_d^2/(Q^2)}},\label{bx}\eeq where $C$ is a
proportionality constant.  For later use we define $\delta
\omega=\omega_r^2 - \omega_d^2$.

The coupled system is held in this state for time, $\tau_{ESR}$,
until steady state has been achieved (B$\rightarrow$C). After
manipulation, rapid adiabatic passage is used to change $\epsilon$
from negative to positive values for charge sensing measurement via
the quantum point contact, QPC (C$\rightarrow$D). The
$|(1,1)T_+\rangle$ state remains in the spin-blocked configuration
whereas the $\left|\uparrow\downarrow\right\rangle$ tunnels directly
to $|(0,2)S\rangle$. A schematic of this pulse sequence is shown in
fig.~\ref{fig2}b. After manipulation, the spin state is a mixture of
$|(0,2)S\rangle$ and $|(1,1)T_+\rangle$ states (at point D). In this
condition, the QPC current, $I_{QPC}$, will be a mixture of the
corresponding (0,2) and (1,1) charge states. At off resonance, the
spin configuration at (D) will be $|(0,2)S\rangle$ with (0,2) QPC
signal. Therefore, by sweeping the tuning field across
resonance, a peak in the $I_{QPC}$ is expected. 

\section{Master equation of ESR Hamiltonian}

\begin{figure}[!t]
\includegraphics[width=1\columnwidth]{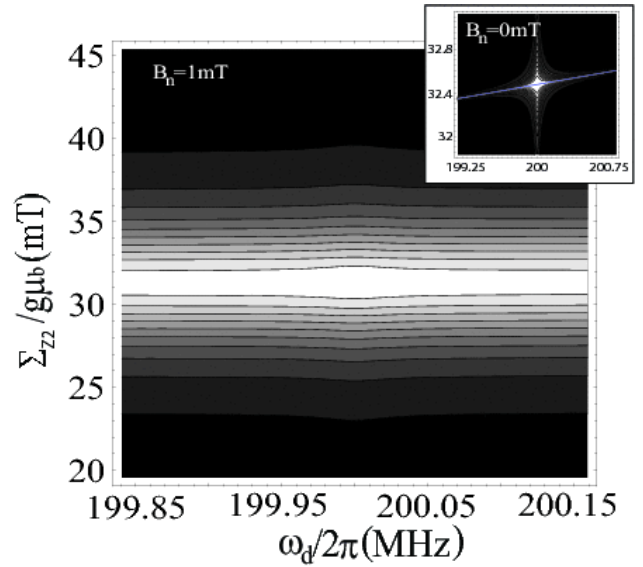} 
\vspace{-.5cm} \caption{\label{fig4} A contour plot of the expected
QPC measurement for the occupation of the $\ket{1,1}T_+$ state as a
function of the field, $\Sigma_{z2}/g\mu_B$, and the driving
frequency of the resonator $\omega_d$, for long ESR mixing time,
$\tau_{ESR}$, and hyperfine interaction $\ex{B_{N}}=1mT$. The inset
shows the same response for the zero hyperfine interaction.  The
horizontal arm of the `star' is defined by $\Sigma_{z2}/\hbar -
\omega_d=0$, as illustrated by the superimposed line in the inset.
When the hyperfine interaction is on, the horizontal arm is
broadened considerably, with only a small exaggerated response at
the $\omega_r=\omega_d$ point. Here, white is maximum ($0.42$ for
the main figure, $\approx 0.5$ for the inset), black is minimum.}
\end{figure}

The response of the QPC can be modeled using a master equation model
for $H_{ESR}$. To robustly model the coupled system, the various
mechanisms by which the electron spins can decohere are considered.
The $|(0,2)S\rangle$ is initialized at point A in the schematic of
the pulse sequence shown in fig.~\ref{fig2}b. On experimental time
scales, this state is highly robust and no spin dephasing takes
place. Concurrently the mechanical element is at resonance and
thermal fluctuations in its motion can result in the ESR field
smearing out i.e. the time dependence of the ESR field shown in the
inset of fig.~\ref{fig1}b will result in small amplitude
oscillations being super-imposed on the sinusoidal ESR response.
However, the amplitude of these thermal fluctuations can readily be
extracted from the equipartition theorem and for the operating
temperature of this experiment ($\sim$50 mK) the thermal amplitude
is expected to be $\sim$0.1 pm which is more than 4 orders of
magnitude less than the amplitude of the beam. Therefore, spin
dephasing due to thermal noise in the mechanical element is
negligible. Furthermore, the Curie temperature of the Co
micro-magnet is 1394 K which is much greater than the operating
temperature of the proposed experiment. Hence spin dephasing due to
fluctuations in the ferromagnetic properties of Co are again
negligible. In general electrical noise (the detuning parameter,
$\epsilon$, is varied by pulsing the voltages of the QD gates)
induced fluctuations in the exchange energy, $J$ (via $\Delta$ and
$T$ noise) lead to extra mixing between the $|(0,2)S\rangle$,
$|S\rangle$ and $|T_0\rangle$ states. However, because the spins are
manipulated at large detuning, gate voltage noise can be neglected
because $dJ/d\Delta \sim$0 in this regime, i.e. the states are to
first order insensitive to electrical noise. As demonstrated in ref.
\cite{spinorbit}, spin-orbit coupling, which is an extra decoherence
channel can be neglected in GaAs quantum dot system for the proposed
experimental timescales. The remaining main decoherence mechanism
for the electron spin is the Hyperfine interaction with nuclear
spins within the quantum dots \cite{nuclei}.  We ignore the
hyperfine interaction for the electron in the left dot by assuming
the mean nuclear magnetic field is much smaller than the Zeeman
splitting, thus the spin state is not flipped. In addition, because
of the large energy difference between the two dots, we can ignore
inelastic singlet-singlet transitions.

Under these conditions, the master equation describing the spin states of the electron in
the right dot is,

\beq \hbar\frac{d \rho(t)}{dt} &=& -i [H_{ESR}+H_N,\rho(t)]
+L[\rho(t)], \eeq

where $H_N=-g.\mu_B(B_N.\sigma_2)$ is the hyperfine interaction and
$L[\rho(t)]$ describes the incoherent Markovian spin relaxation at
rate $\gamma$.  We reach the steady state solution by taking
$t\rightarrow \infty$, $\frac{d \rho(t)}{dt}=0$, and thus inverting
to find $\rho(t\rightarrow \infty)=[-i\hbar [H_{ESR}+H_N,\rho(t)]
+L]^{-1}$. Assuming the rotating wave approximation, we are able to
obtain an analytical solution for the
occupation 
of the $\ket{(1,1)T+}$ and $\ket{(0,2)S}$ states.


For static hyperfine nuclear fields $B_N$, the steady state of the
density matrix elements for the diagonal $\ket{(1,1)T_+}$,
$\ket{(0,2)S}$, and the off-diagonal $\bra{(1,1)T_+}\rho
\ket{(0,2)S}$, states are (omitting the notation $t\rightarrow
\infty$),

\begin{widetext}
\beq \rho_{(1,1)T_+} &=& \frac{16B_{Y,N}^2 + (4B_{X,N} +
\Sigma_{x2})^2}{32(B_{X,N}^2 + B_{Y,N}^2 +2B_{Z,N}^2) +
32\Sigma_{x2} B_{X,N} +64 \hbar B_{Z,N} \Delta \omega +
2\Sigma_{x2}^2 + \hbar^2 16\Delta \omega^2 + 4\hbar^2\gamma^2}\\
\rho_{(0,2)S} &=&
1-\rho_{(1,1)T_+} \\
\rho_{(1,1)T_+,(0,2)S} &=& \frac{(4(B_{X,N} - iB_{Y,N}) +
\Sigma_{x2})(4B_{Z,N}+2\hbar\Delta\omega-i\hbar\gamma)}{16(B_{X,N}^2
+ B_{Y,N}^2 +2B_{Z,N}^2) + 16\Sigma_{x2} B_{X,N} +32 \hbar B_{Z,N}
\Delta \omega + \Sigma_{x2}^2 + \hbar^2 8\Delta \omega^2 +
2\hbar^2\gamma^2}. \eeq
\end{widetext}
where $\Delta \omega= \Sigma_{z2}/\hbar - \omega_d$ is the Rabi
frequency, $\gamma$ is a phenomenological damping of the spin state,
and $B_{i,N}$ is the nuclear field in the $i$th direction.  The
off-diagonal term is in the rotating frame.  We assume that the
nuclear field is static on the $\hbar/\Sigma_{x2}$ and $1/\gamma_1$
time scales, but is fast on the time scale of collating a clear QPC
measurement. Thus, in the data presented in fig.~\ref{fig3}, we
integrate over a normal distribution of possible nuclear spin
orientations\cite{christo} with mean value
$\ex{B_{N}}=\sqrt{3}\ex{B_{i,N}}$, and a standard deviation of
$\sqrt{2}\ex{B_{N}}$, so \beq
\ex{\rho_{(1,1)T_+}}_{B_N}&=&\nonumber\\ \int_{-\infty}^{\infty}
\rho_{(1,1)T_+}&\prod&_{i=x,y,z} \frac{e^{\left[\frac{-(B_{i,N}^2
-\ex{B_{i,N}^2})^2}{4\ex{B_{i,N}}^2}\right]}}{(2\ex{B_{i,N}}\sqrt{\pi})}
\text{d}B_{i,N}.\eeq Consequently, $\ex{B_{N}}$ is the important
quantity in determining the affect of the nuclear field on the ESR
measurement. In fig.~\ref{fig3}, the effects of the hyperfine
coupling on the QPC measurement are shown. For $\ex{B_{N}} <
\Sigma_{x2}$ the on resonant response of the QPC is clearly visible.
For $\ex{B_{0,N}\geq \Sigma_{x2}}$ the on resonance peak becomes
smeared out and eventually obliterated.


\begin{figure}[!t]
\includegraphics[width=1\columnwidth]{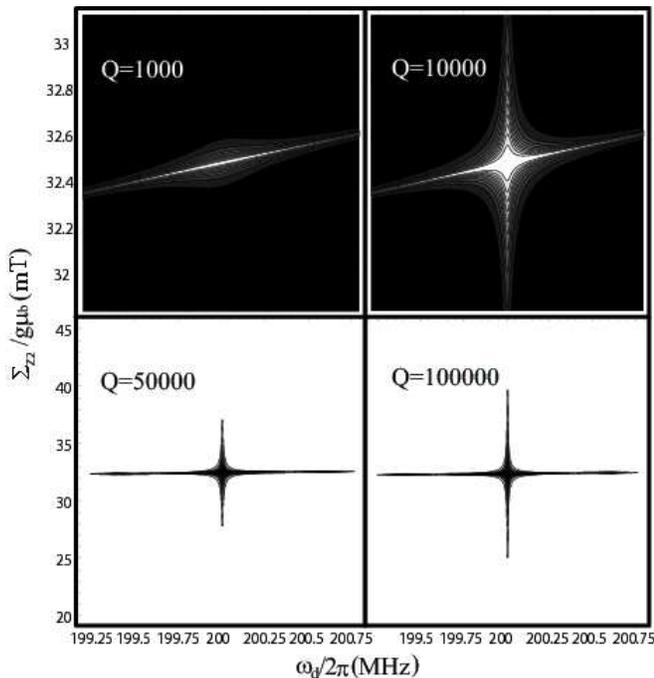}
\vspace{-.5cm} \caption{\label{fig5} Contour plots of the expected
QPC measurement for the occupation of the $\ket{1,1}T_+$ state as a
function of the field, $\Sigma_{z2}/g\mu_B$, and the driving
frequency of the resonator $\omega_d$, for zero hyperfine
interaction, and a range of $Q$ factors.  Changing the Q factor
alters the range of the vertical arm ($\omega_d - \omega_r=0$) of
the star.  In the top two figures, white is maximum.  In the bottom
two figures, black is maximum (for clarity).}
\end{figure}

\section{ESR line width and resonator quality factor}

Ignoring the hyperfine interaction and the stochastic decoherence
rate of the spin ($\gamma=0$), we can see very clearly how the line
width of the ESR response depends on the Quality factor of the
resonator.  The ESR response is,
 \beq
\rho_{(1,1)T_+}=\frac{1}{2(1+\frac{2\hbar^2}{\Sigma_{x2}^2} \Delta
\omega^2)}. \eeq However, as mentioned earlier, the field
$\Sigma_{x2}$ is both $Q$ and $\omega_d$ dependant.  This gives the
following Lorentzian form with two maxima,
 \beq
\rho_{(1,1)T_+}=\frac{1}{2(1+\frac{32\hbar^2 \omega_d^2
\omega_r^2}{Q^2 C^2} \Delta \omega^2 + \frac{8 \hbar^2 \Delta
\omega^2}{C^2} \delta \omega^2)}. \eeq Thus the line width of the
ESR response, obtained by sweeping the tuning field, is \beq
\gamma^{ESR}=\frac{QC}{\hbar\sqrt{8(4\omega_d^2\omega_r^2 +
Q^2(\omega_d^2-\omega_r^2)^2)}}. \eeq  where $C$ again is the
constant defined by properties of the resonator (see Eq.
[\ref{bx}]). When the resonator is driven at resonance
$\omega_d=\omega_r$, the ESR line width is $\gamma_L\approx Q$, the
inverse of the resonators own line width, Eq. [\ref{bx}]. This can
be understood by the fact that the reduced motion of the resonator,
because of a lower Quality factor, implies a reduced maximum
$\Sigma_{x2}$ field, which in turn reduces the steady state
probability for the spin to be in the $\ket{(1,1)T_+}$ state when
the Rabi tuning $\Delta \omega$ is off-resonance.  Fig.~\ref{fig5}
illustrates the effect of changing $Q$ on the QPC measurement.

In addition, for large detuning of the resonator driving frequency
from the natural frequency $\delta \omega=\omega_d^2-\omega_r^2$, we
find that $\gamma_L\approx 1/\sqrt{(\omega_d^2-\omega_r^2)}$. Hence
the $\Delta \omega$ peak becomes very sharply defined and almost
independent of the quality factor.  Thus the transduction of the
resonator motion by ESR is advantageous because it allows us to read
out both the quality factor $Q$ and intrinsic properties of the
resonator via the constant $C$.  Furthermore, for large detuning
$\omega_d^2-\omega_r^2$, the spin rotation can be performed with a
very narrow line width.

Alternatively, by fixing the tuning field at $B_0 = \hbar
\omega_r/g\mu_B - B_{z2}$ ($B_{z2}$ can be measured in principle by
DD spectroscopy \cite{DDspec}) and sweeping the frequency of the
mechanical oscillator about the resonance point, $\delta
\omega=\Delta \omega=0$, a peak in $I_{QPC}$
is again seen.  But now the corresponding line shape 
is the convoluted response of the nano-mechanical oscillator
and the two-electron system. 
If the tuning field is is moved away from the $\omega_r/g\mu_B$
point, both maxima in the double Lorentzian are visible. In this
case, the two peaks both have a $\omega_d$ dependant line shape,
which dominates over the $Q$ dependance.



Additionally, moving beyond the steady state, the mechanical
resonator provides a local ESR field which enables the electron spin
states to be manipulated. For example, by modulating the waiting
time (B$\rightarrow$C) to provide a $\pi$ pulse so that, $\tau_{ESR}
= \tau_{\pi} = 2\pi\hbar / \Sigma_{x2}$ = 0.2 $\mu$s , then the
final state at readout is the triplet $\left|T_{+}\right\rangle$. By
employing a larger tuning field,  the states
$\left|\uparrow\downarrow\right\rangle$ and $|T_-\rangle$ can also
be resonantly coupled so that the spin of the left dot is flipped
given the tuning field is less than the Co switching field so that
the magnetisation does not reverse.

In the above described coupled system, actuation of the mechanical element is achieved via the piezoelectric effect between the top gate defined over one of the clamping points of the beam and the shallow buried 2DEG \cite{Imran}. The mechanical resonance is then excited by applying an AC voltage to the top gate at the mechanical resonance frequency. Crucial to the success of this scheme is the in-plane alignment of the micro-magnet and the shallow buried 2DEG on the mesa on which the DQDs are defined. This can be achieved by applying a DC voltage to the top gate on the mechanical element which will induce a static deflection of the mechanical beam enabling alignment to be achieved as visualised in the inset of fig.~\ref{fig1}a. Simultaneously a small AC modulation will drive the beam to mechanical resonance.
\section{Conclusions}

A two-level system realized in a two-electron double quantum dot is
a highly sensitive quantum detector. Magnetically coupling the
double quantum dots to a high frequency nano-mechanical resonator
will allow the displacements of the resonator to be detected. Tuning
the Larmor frequency to match the resonators natural frequency, the
electron spin in one of the dots can be flipped. By simultaneously
measuring the charge state of the double quantum dots, this
transition can be detected thus enabling the natural frequency of
the mechanical resonator to be determined.

The authors NL, IM and MP-L have contributed equally to this
manuscript. The authors also acknowledge A. Blais for useful
discussions and for critically reading the manuscript. S. Camou, S.
Etaki, C. Buizert and H. Inagaki are thanked for their help and
encouragement. We acknowledge financial support from the
Grant-in-Aid for Scientific Research A (No. 40302799), the Special
Coordination Funds for Promoting Science and Technology, MEXT,
SORST-JST, DARPA grant no. DAAD19-01-1-0659 of the QuIST program and
JSPS KAKENHI(16206003) and JSPS grant 17-05761.

\bibliography{bibliography}

\end{document}